\documentclass[a4paper,twocolumn,prl,floatfix,aps,10pt]{revtex4-1}
\usepackage{textgreek}

\usepackage{amsfonts,amsmath,amssymb}
\usepackage{textcomp}
\usepackage{bm}
\usepackage{graphicx}
\usepackage[latin1]{inputenc}
\usepackage{rotating}
\usepackage{color}
\usepackage{hyperref}
\usepackage{siunitx}
\usepackage{natbib}

\newcommand{\beq}[1]{\begin{equation} \eqlab{#1}}
\newcommand{\eeq}{\end{equation}}
\newcommand{\bsub}{\begin{subequations}}
\newcommand{\esub}{\end{subequations}}
\def\bal#1\eal{\begin{align}#1\end{align}}
\def\bsubal#1\esubal{\bsub \begin{align}#1\end{align} \esub}

\newcommand{\eqlab}[1]{\label{eq:#1}}
\renewcommand{\eqref}[1]{Eq.~(\ref{eq:#1})}

\begin{document}

\date{\today}

\title{Solutal Marangoni flow as the cause of ring stains from drying salty colloidal drops }

\author{ Alvaro Marin$^{*\mathrm{(a)}}$, 
Stefan Karpitschka$^{*\mathrm{(b)}}$
Diego Noguera-Mar\'in$^{\mathrm{(c)}}$,
Miguel A. Cabrerizo-V\'ilchez$^{\mathrm{(c)}}$,
Massimiliano Rossi$^{\mathrm{(d)}}$, 
Christian J. K\"ahler$^{\mathrm{(d)}}$,
and
Miguel A. Rodr\'iguez Valverde$^{\mathrm{(c)}}$}

\affiliation{
$^{\mathrm{(a)}}$Max-Planck Center for Complex Fluid Dynamics, University of Twente, The Netherlands\\
$^{\mathrm{(b)}}$Max-Planck Institute for Dynamics and Self-Organization (MPIDS), D-37077 G\"ottingen, Germany\\
$^{\mathrm{(c)}}$Biocolloid and Fluid Physics Group, University of Granada, Spain\\
$^{\mathrm{(d)}}$Institut f\"ur Str\"omungsmechanik und Aerodynamik, Universit\"at der Bundeswehr M\"unchen, Germany\\}

\begin{abstract}

Evaporating salty droplets are ubiquitous in nature, in our home and in the laboratory. Interestingly, the transport processes in such apparently simple systems differ strongly from evaporating ``freshwater'' droplets since convection is partly inverted due to Marangoni stresses. Such an effect has crucial consequences to the salt crystallization process and to the deposits left behind. In this work we show unprecedented measurements that, not only confirm clearly the patterns of the flow inversion, but also elucidate their impact on the distribution of non-volatile solutes. Contrary to what has been often reported in the literature, such a flow reversal does not prevent the formation of ring-shaped stains: particles accumulate at the contact line driven solely by the interfacial flow. We can therefore conclude that the classical ``coffee-stain effect'' is not the only mechanism that can generate ring-shaped stains in evaporating droplets. 

\pacs{\\\\$^*$ Corresponding author: a.marin@utwente.nl}
\end{abstract}

\maketitle


Evaporating sessile liquid droplets are ubiquitous in our daily enviroment and in many industrial applications, for instance in \emph{evaporative self-assembly} techniques \cite{Wei2012coffeerings}. Consequently, evaporating sessile droplets have been intensively studied recently, in particular after the seminal work of Deegan et al.\cite{Deegan:1997vb}. In most applications, the droplet does not only contain a dispersion of deposit particles in a pure solvent, but also many other, soluble components. If such components are surface-active, the transport processes inside the droplet can be drastically affected. That has been shown for bulk mixtures \cite{karpitschka2017contraction, karpitschka2012noncoalescence, Cira:2015ii,Yaxing:2018} and surfactant solutions \cite{Marin:2016surfactant, Sempels:2013bi,Bruning:2018Coalescence}.

The effect of non-volatile solutes, like salts, is paramount since they are part of most liquids in biomedical applications to match the ionic strength and the pH of the human body. Salty droplets are of particular interest also in the pharmaceutical industry where controlled crystalization is crucial for drug formulation \cite{Zheng:2003baa}. Salt solutions in contact with metallic surfaces provoque corrosion, and although ubiquitous, the underlying mechanisms remain poorly understood \cite{li2015one, Soulie:2015}. 
In most cases of salty droplet evaporation, the residual components enrich near the contact line. Therefore, crystallization and growth of the final deposit deposit typically nucleates at the contact line \cite{ShahidzadehBonn:2008br,Shahidzadeh:2015cr,Soulie:2015}.
Recent artwork of the photographer Maurice Mikkers displayed micrographs of dried teardrops, itemized by the emotional state under which the tears were shed \footnotetext[1]{http://www.imaginariumoftears.com}\cite{Note1}.

\begin{figure}[b!]
  \centering
 
\includegraphics[width=\columnwidth]{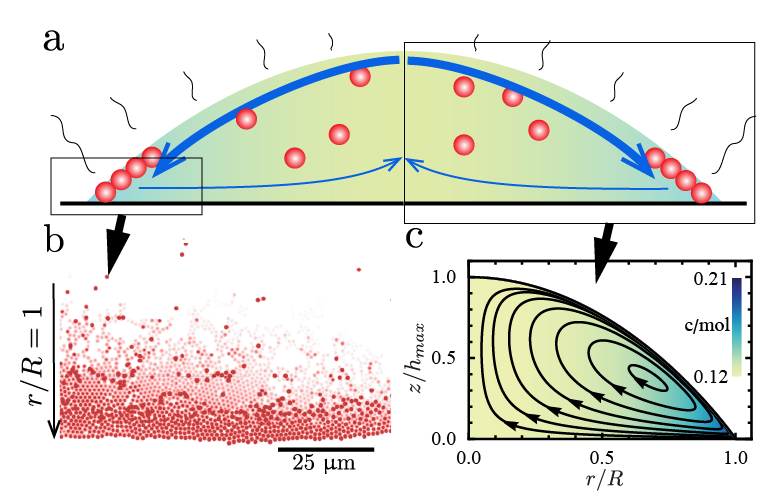}
\caption{(a) Sketch of a sessile droplet containing a non-saturated aqueous saline solution and a dilute dispersion of polymer microparticles. The blue lines illustrate the direction of the flow within the droplet. Note that a large portion of the liquid (close to the substrate) flows in opposite direction as in the classical ``coffee-stain effect''. (b) Particles reach the contact line along the liquid-gas interface, forming a ring-shaped stain. (c) The inverted vortical flow profile is shown through the computed streamlines, which is also measured experimentally via particle tracking velocimetry. \label{fig:1}}
\end{figure}

\begin{figure*}
\centering
\includegraphics[width=.85\columnwidth]{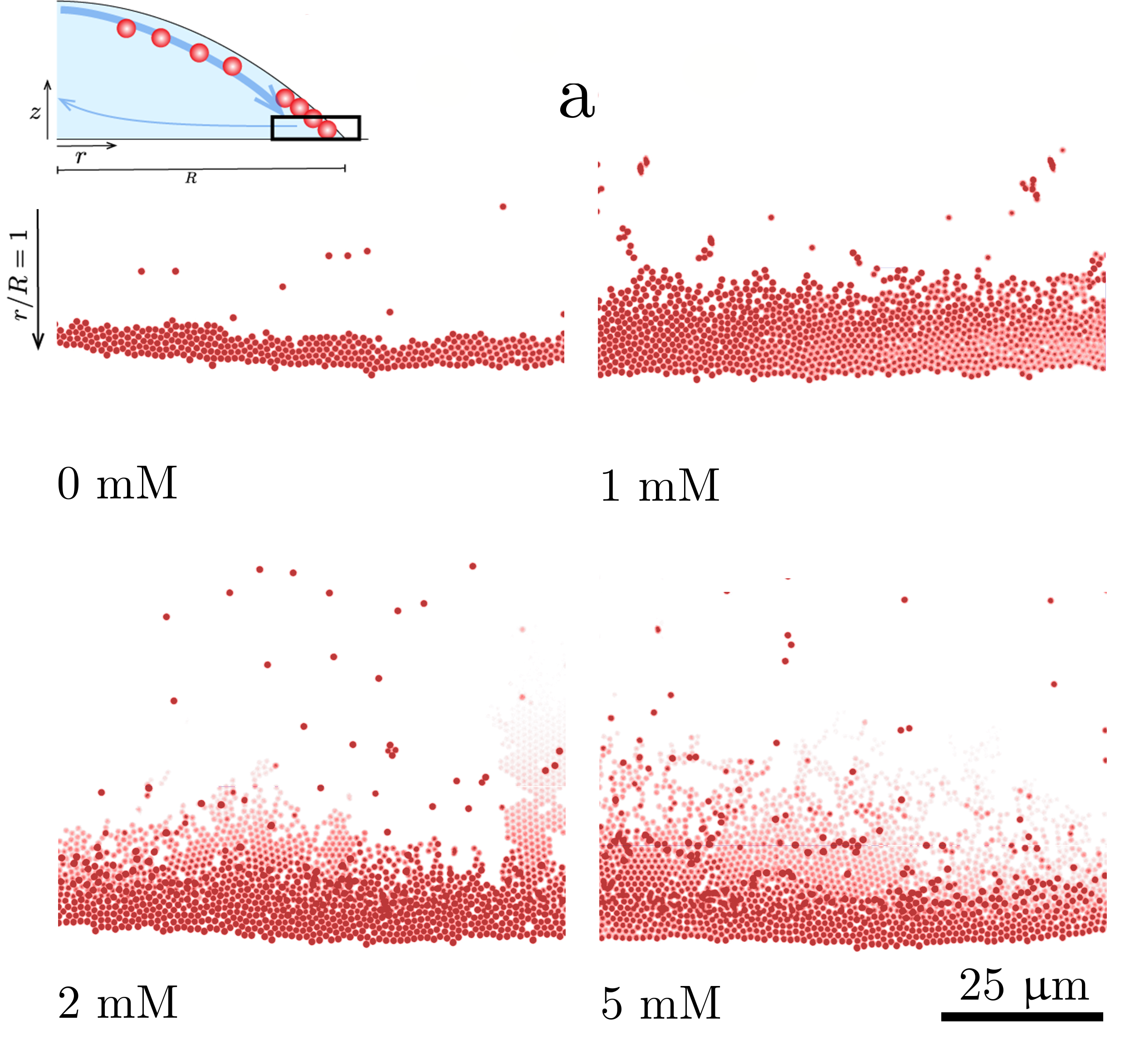} \includegraphics[width=.85\columnwidth]{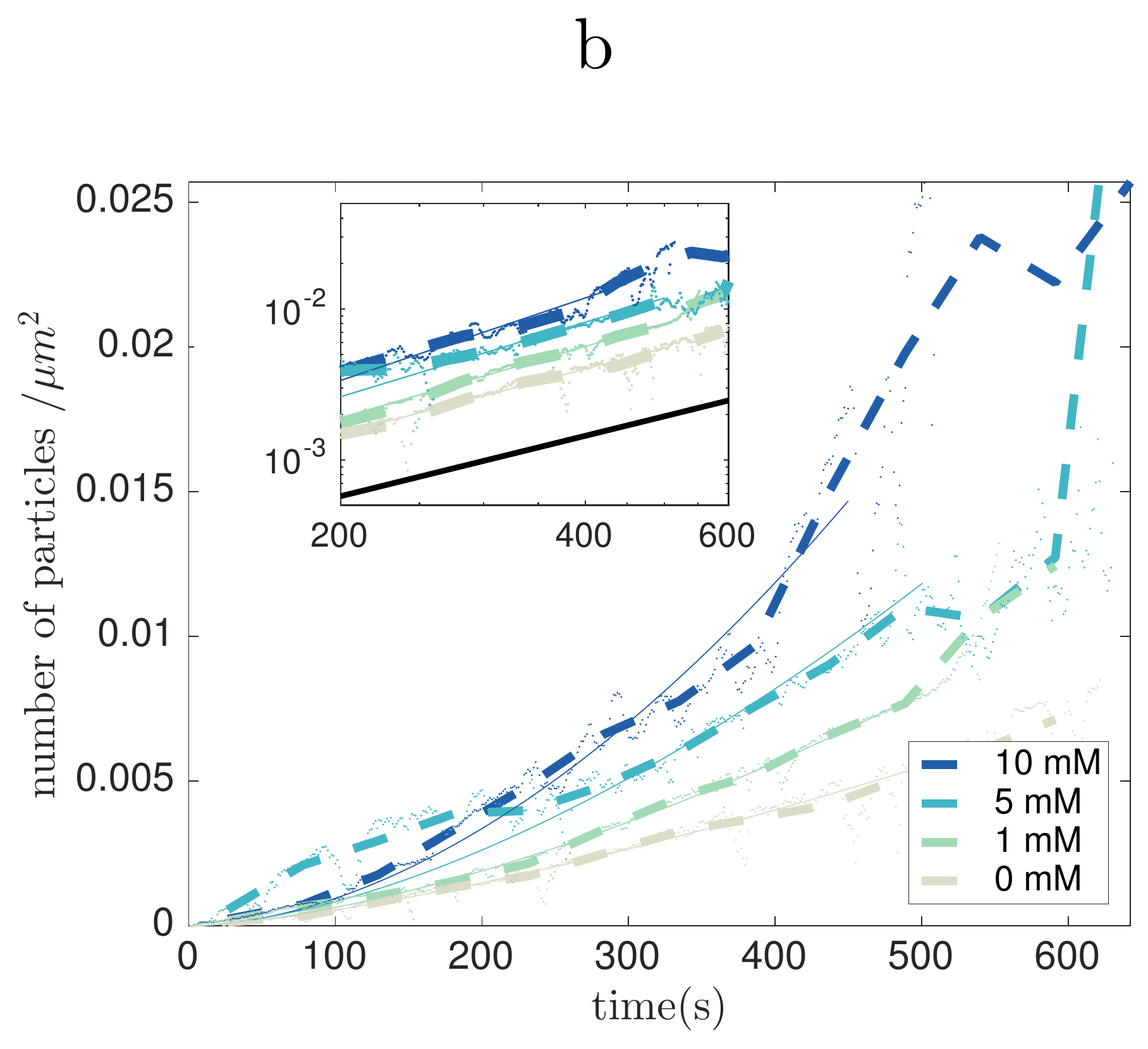}
\caption{(a) Images close to the contact line depicting the particle accumulation in the final stage of drop evaporation for increasing NaCl concentrations. (b) Number of particles near the contact line per unit of wet area against the time for different solute concentration. The dashed lines show the binned data. Inset: Final values of normalized particle increment in terms of NaCl concentration. Black continuous line corresponds to the stain growth $N\propto t^{1.33}$, reported by Deegan et al. \cite{Deegan:1997vb} for comparison. \label{fig:confocal}}
\end{figure*}

Peripheral salt enrichment leads to Marangoni convection, which in turn affects the solute distribution~\cite{Kim:PRL2016}. Chaotropic salts act as ``anti-surfactants''\cite{wilson:antisurfactants} and increase the surface tension of water. Thus the associated Marangoni convection is directed toward the contact line. Under pinned contact line conditions, volume conservation then leads to a reversal of the capillary flow near the substrate, which is opposite to the case of pure water. The existence of such a nonmonotonic radial flow patterns provokes a simple yet non-trivial question: Will the deposits be ring-shaped as in the classical \emph{coffee-stain} effect? 

In this paper we present experimental and numerical evidence for solute-induced Marangoni stresses that dominate the whole flow within an evaporating droplet. We also demonstrate the existence of a new type of ring-shaped stain that, due to a surface-driven Marangoni flow, assembles along the liquid-vapor interface. This is in stark contrast to the commmonly observed assembly in bulk for ``freshwater'' droplets \cite{Deegan:1997vb}. To show this unambiguously we first observe and quantify the growth of the particle deposits in evaporating water droplets for different salt concentrations using confocal imaging. Secondly, we measure the full three-dimensional flow inside the evaporating salty droplet using 3D particle tracking velocimetry. Finally, we compare the experimental results with numerical simulations that capture the solvent evaporation, the evaporation-induced liquid flow and the quasi-equilibrium liquid-gas interface.

\begin{figure*}[t]
  \centering
\includegraphics[width=2\columnwidth]{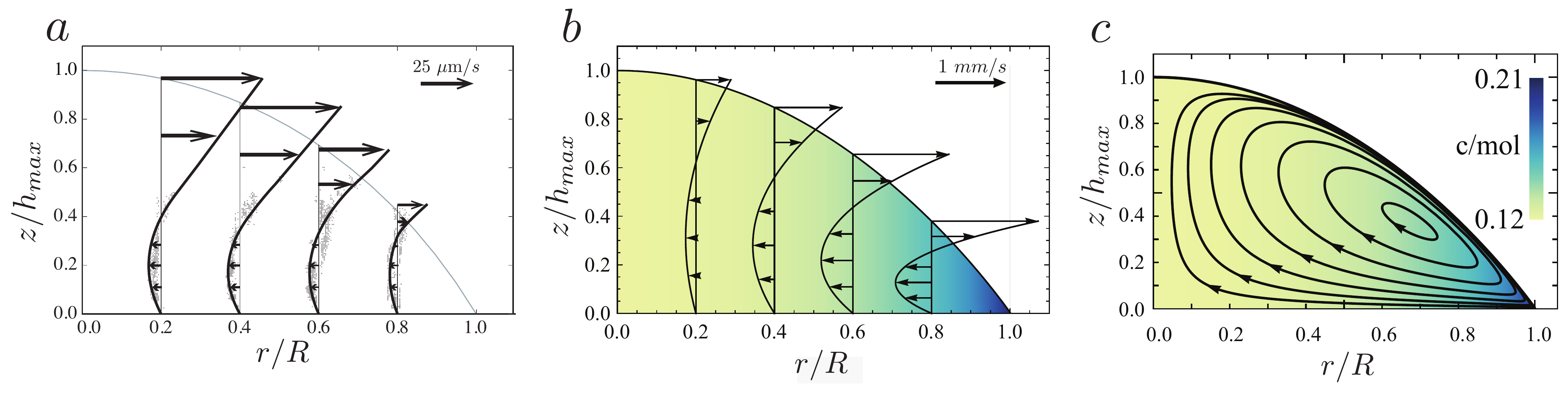} 
\caption{(a) Experimentally obtained flow profile in an evaporating droplet obtained by a 3D particle tracking (APTV \cite{Cierpka2011,rossi2014optimization}).
(b) Velocity profile and (c) flow streamlines in an evaporating droplet with an initial NaCl concentration of 100 mM obtained by solving the Stokes flow inside the droplet in the lubrication approximation. The instant shown corresponds to 50\% of the total evaporation time.
 The experimental conditions are identical as those in the simulations shown in (b) and (c). Note that in both simulations and experiments the interfacial flow towards the contact line dominates the flow in the droplet \label{fig:comparison}}
\end{figure*}

The system that we are considering is summarized in Fig. \ref{fig:1}: a sessile water droplet containing a sodium chloride (NaCl) solution is deposited on a glass slide and evaporates \textcolor{black}{at room temperature and an ambient relative humidity of $40\,\%$}. We used initial concentrations of $C_o=$ 0, 1, 2, 5 and 10 mM, far from saturation. In order to study how the deposit structure is formed, we use confocal microscopy to observe a thin optical plane close to the contact line and to the glass substrate. A small amount of charged-stabilized polystyrene fluorescent particles are introduced in the droplet (particle diameter 1.11 $\mathrm{\mu m}$, Microparticles GmbH, below 0.001~\% w/w)\footnotetext[2]{See Supplemental Material at [URL will be inserted by publisher] for more details on the experimental conditions and on the Astigmatism Particle Tracking Velocimetry method.}\cite{Note2}.

Figure \ref{fig:confocal}(a) shows snapshots of the contact line region in the last stage of solvent evaporation, \textcolor{black}{at identical times after droplet deposition}. The number of initially dispersed particles was kept constant for all experiments, only the initial salt concentration $C_o$ was increased. 
Unambiguously, the amount of particles that aggregate at the contact line increases with $C_o$. On panel (b), we quantified the number of particle per unit of wet area against time. In a $5$~mM salt solution, aggregation is accelerated by a factor of 2 as compared to pure water.
The increase of particle flux with the amount of salt can be further quantified by fitting a simple power law to the amount of particles per unit area $N\sim t^{\alpha}$. Accordingly, we obtain exponents that increase with the amount of salt in the system, from $\alpha \sim 1.4$ for 0 mM, which compares well with the results reported by Deegan et al. \cite{deegan2000contact}, up to $1.8$ for 10 mM (inset in Fig. \ref{fig:confocal}b). The mechanism of this enhanced ring-stain formation is revealed by the shading and the defocusing of the particles in Fig. \ref{fig:confocal}(a): lighter tones correspond to particles that are more distant from the focal plane and thus from the substrate. At high salt concentrations ($C_o\ge2$~mM), most of the particles reach the contact line while following the water-air interface, driven by the solutal Marangoni flow due to salt enrichment. As the initial salt concentration $C_o$ is increased further, well-ordered structures of defocused particles come into sight, i.e. at the liquid-air interface. At the same time, fewer particles appear in the focal plane, i.e. just above the liquid-substrate. 

The observation of the deposit formation yields two surprising effects: (1) an increase in the particle flux with the amount of salt and (2) the arrival of the particles mostly along the interface and not through the bulk, as is it typically observed in `freshwater' droplets. 

Since an explanation of this effect must be deeply connected to the convection in the droplet, we next measure the three-dimensional velocity field using Astigmatism Particle Tracking Velocimetry (APTV) \cite{Cierpka2011,rossi2014optimization}. This is a single-camera technique that yields all velocity components for each tracer particle in space over time \cite{Note2}. The experimental results are shown in Fig. \ref{fig:comparison}a \footnotetext[3]{Fig. S1, in which particle trajectories from the measurements are expliticily shown.}\cite{Note3}, which clearly shows the ``inverted'' flow profile inside an evaporating water saline droplet. It consists of a vortical flow structure that drives the liquid in the bulk toward the droplet's center, and the liquid close to the liquid-air interface toward the contact line. Consequently, most of the particles reach the contact line  through the droplet's surface, and not through the bulk as it happens with pure water solutions and in the classical \emph{coffee-stain} effect.

Contrary to surfactant solutions, chaotropic electrolytes increase the surface tensions of water, as it is well known from classical thermodynamics \cite{Debye1923, onsager1934surface, levin2001surface}. For NaCl, this effect quantifies as $\sim1.5$~mN/m per mol NaCl per liter of water \cite{matubayasi1999thermodynamic}. Therefore, it is expected that an accumulation of salt at the contact line would generate a surface tension gradient toward the droplet's border. The gradient is clearly strong enough to overcome the evaporation-driven bulk flow or any additional thermal Marangoni convection. Due to the pinning condition at the contact line, the outward directed Marangoni flow must be compensated by an ``inverted'' capillary flow toward the center of the droplet.

To get a deeper insight into the evaporation process, the resultant flow, and the possible particle accumulation mechanisms, we performed numerical simulations for the advection-diffusion problem in the droplet in the lubrication approximation on a radially symmetric domain. The fluid flow is coupled to the diffusion problem in the gas-phase via Raoult's law and a Stefan condition. The salt is introduced through a compositional variable that diffuses, is advected with the bulk flow, and alters the physical properties of the liquid. Most importantly, it alters surface tension and vapor pressure, for which we have used literature values~\cite{ZhiBao1999,Partanen2009,RubberBibble}. One important limitation of the model is that crystallization is not explicitly modeled. Although saturation conditions are not reached in the simulations, this is typically the case during the latest stage of the experiments, \textcolor{black}{but only after the initial particle deposit structure could be observed.}

\begin{figure*}[t]
\includegraphics[width=1.85\columnwidth]{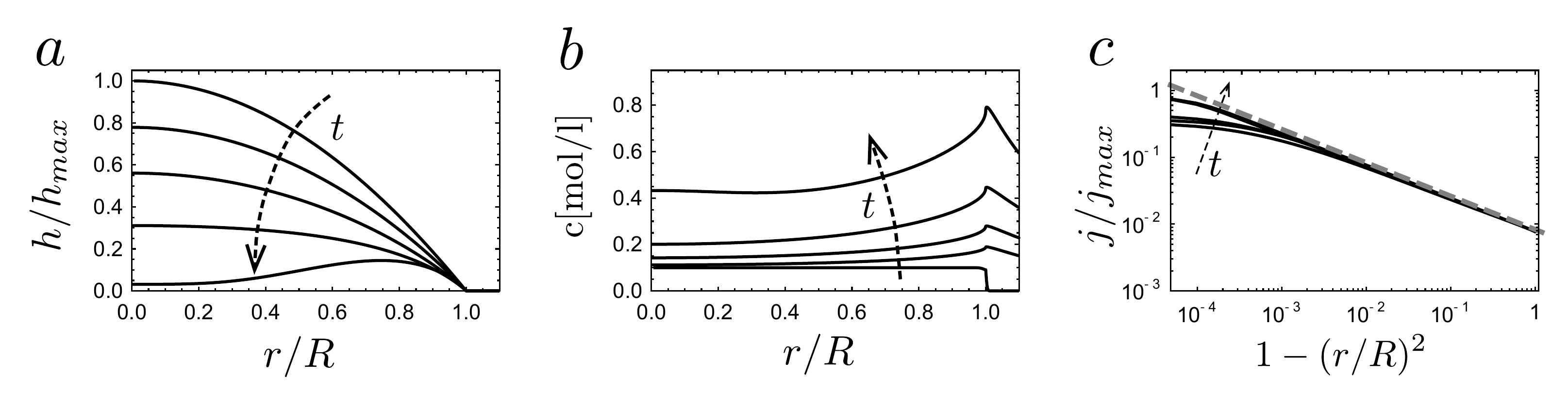}

\caption{Results from the numerical simulations on the evaporating salty droplets. The plots correspond to different stages in the evaporation of a droplet containing NaCl at a concentration of 100 mM, initial contact angle of $20^\circ$ and radius of 1 mm. The different lines correspond to 0, 20, 40, 60 and 80\% of the total evaporation time:(a) droplet interface h(r) in time, (b) salt concentration $c$ in mol/l (c) and normalized evaporative flux $j/j_{max}$, which becomes singular as the contact line is approached ($r/R\rightarrow 1$)\cite{Deegan:1997vb}. \label{fig:sims}}
\end{figure*}

The results of the numerical simulations are shown in Fig. \ref{fig:comparison} and \ref{fig:sims}. In Figure \ref{fig:comparison}(b) and \ref{fig:comparison}(c) it can be seen that the quality of the resultant flow structure is identical as compared to the experiments (Fig. \ref{fig:comparison}(a)). The simulations show a strong accumulation of salt at the contact line (Fig. \ref{fig:sims}(b)) and consequently, a strong surface tension gradient directed towards the contact line. The surface tension gradient generates the Marangoni flow observed, which dominates the flow within the droplet (Fig. \ref{fig:comparison}). Note that the singularity in the evaporative flux at the contact line is regularized only at a (sub-)micron length scale for millimetric droplets (Fig. \ref{fig:sims}c) \cite{deegan2000contact,popov2005evaporative}. Thus the evaporation rate peaks sharply at the contact line, and peak height and width change by only about half a decade over time, due to salt enrichment. 

Despite the qualitative analogy in the flow profiles computed numerically and measured experimentally, a significant quantitative disagreement is found in the value of the velocities (explicit in Figure \ref{fig:comparison}). Simulations predict much stronger Marangoni convection (see Fig.~\ref{fig:comparison}). Simulations have been repeated using diferent methods (Finite elements and finite volume schemes, using various software packages) to exclude technical issues like the level of approximation or discretization effects \footnotetext[4]{Private communication with Christian Diddens}\cite{Note4}. Experiments have been replicated in different laboratories, with different equipment and samples, always obtaining the same results. It is interesting to note that a similar quantitative disagreement is found in water droplets in thermal Marangoni flows \cite{Massi2018Water} and with certain bi-component systems \cite{diddens2017detailed}. Given the consistency of the experimental results despite different preparation methods, and although we are lacking a physical explanation for such a disagreement, we do not invoke here the commonly accepted explanation through surface active trace impurities~\cite{Hu:2005dv}. Rather, these results point toward an unknown physical phenomenon that is not properly captured by our current understanding of continuum models, and is thus absent in the simulations. Nonetheless, the simulations confirm the presence of the Marangoni flow with the same structure found in the experiments.

From the information gained in experiments and simulations, we can conclude that particles arriving at the contact line for salt concentrations above 2 mM are either adsorbed at the liquid-air interface or stabilized very close to it, forming an interfacial monolayer of particles that is advected towards the contact line by Marangoni flow. The increase of initial salt concentration (within the range 0-200 mM) also increases the maximum concentration gradient achievable, increasing the interfacial velocity and therefore the flux of particles towards the contact line. 

The accumulation of particles at the liquid-air interface is crucial in this process. Particles that follow streamlines which intersect the moving interface tend to remain there and follow the interfacial flow. This phenomenon has been reported using different three-dimensional visualization techniques \cite{trantum2013cross} and it is also clearly observed using our APTV measurements~\cite{Note2}. Recently, Kang et al. \cite{Kang:2016} proposed a mechanism for the formation of ring-shaped stains in water droplets solely based on the capture of particles at the receding interface and subsequently transported along the interface until they are deposited near the contact line. The receding motion of a contact line causes -- in the laboratory frame -- a receding surface flow of the same magnitude~\cite{Eggers2004}. In addition, sessile water droplets evaporating in atmospheric conditions experience a weak but persistent thermal Marangoni flow which is directed away from the contact line \cite{Marin:2016surfactant}. Thus, particles close to the surface are driven away from the contact line, and it is known that Marangoni flows of such orientation may prevent ring-stain formation~\cite{Kim:PRL2016}. Evaporating droplets of particles dispersed in pure water always show ring-shaped stains, despite such an adverse interfacial flow. This means that particles must reach the stain through the bulk, and that the mechanism proposed by Kang et al. cannot properly capture the physics of ring stain formation. However, the interfacial accumulation mechanism proposed by Kang et al.\cite{Kang:2016} would be valid whenever there is a dominant outward-directed interfacial flow, as in the case of salty droplets. 

In conclusion, we have shown that ring-shaped stains from evaporating ``anti-surfactant'' \cite{wilson:antisurfactants} droplets do not form via the classical, capillary flow based mechanism\cite{Deegan:1997vb}, but via solutal Marangoni flows and interfacial aggregation. This is not only the case for sodium chloride solutions, but similar results have been observed for sodium iodide \cite{Kang:2013kb} and it is expected for several electrolyte solutions. A similar flow inversion has been observed in evaporating ethanol-water mixtures~\cite{Christy2011}, in which the enrichment of water provoques a positive surface tension gradient towards the contact line. This principle also holds for \emph{Ouzo} droplets \cite{Ouzo2016}, however phase separation effects dominate later stages of the evaporation process, causing non-trivial deposit patterns.

Our results suggest that particle arrangements can be tuned from a three-dimensional crystal \cite{Marin:2011hr} to a 2D monolayer by simply modifying the solution's physico-chemical properties.

\begin{acknowledgments}
AM, MR and CK acknowledge financial support by the German Research Foundation grant KA 1808/12-2 and KA 1808/22-1. AM and SK are grateful to Christian Diddens for several discussions on the topic. MARV and MCV acknowledges financial support by the MINECO grant MAT2014-60615-R and MAT2017-82182-R. SK acknowledges financial support from the Max Planck -- University of Twente Center ``Complex Fluid Dynamics -- Fluid Dynamics of Complexity''.
\end{acknowledgments}

\bibliographystyle{apsrev4-1}

%

\end{document}